\documentclass[prl,twocolumn,showpacs,preprintnumbers,amsmath,amssymb]{revtex4}

\usepackage{graphicx}
\usepackage{dcolumn}
\usepackage{bm}

\def\lsim{\mathrel{\lower2.5pt\vbox{\lineskip=0pt\baselineskip=0pt
\hbox{$<$}\hbox{$\sim$}}}}
\def\gsim{\mathrel{\lower2.5pt\vbox{\lineskip=0pt\baselineskip=0pt
\hbox{$>$}\hbox{$\sim$}}}}

\newcommand{\be}{\begin{equation}}
\newcommand{\ee}{\end{equation}}

\newcommand{\NP}[1]{Nucl.\ Phys.\ {#1}}

\newcommand{\PL}[1]{Phys.\ Lett.\ {#1}}

\newcommand{\PR}[1]{Phys.\ Rev.\ {#1}}
\newcommand{\PRL}[1]{Phys.\ Rev.\ Lett.\ {#1}}

\begin{document}

\preprint{hep ph/0202265}

\title{The SU(2) and SU(3) chiral phase transitions 
within Chiral Perturbation Theory}

\author{J. R. Pel\'aez }

\affiliation{Departamento de F{\'\i}sica Te{\'o}rica II,
  Universidad Complutense de Madrid, 28040   Madrid,\ \ Spain\\
and\\ Dipartimento di Fisica. Universita' degli Studi, Firenze,
  Italy and INFN, Sezione di Firenze, Italy}


\begin{abstract}
The SU(2) and SU(3) chiral phase transitions
 in a gas made of pions, kaons and etas
are studied within the framework of Chiral Perturbation Theory.
We describe the temperature dependence of the
quark condensates
by using the meson meson scattering phase shifts
in a second order virial expansion. In particular, the SU(3) formalism yields
a somewhat lower melting temperature for the  non-strange condensates
than within SU(2), and also
predicts that the strange condensate melting is slower
than that of the non-strange, due to the 
different strange and non-strange quark masses.
\end{abstract}

\pacs{11.30.Rd, 11.30.Qc, 12.38.Aw, 12.39.Fe, 11.10.Wx}
\maketitle

There is a growing interest 
in studying the phase diagram of QCD, namely, the 
transition from a hadron gas to a quark gluon
plasma, the chiral phase transitions,
their dependence on the number of flavors, on quark masses,
possible color superconducting states, etc. 

The virial expansion is a simple and successful approach to describe
many thermodynamic features of dilute gases made of interacting pions
\cite{virial2} and other hadrons \cite{virial3}. 
Remarkably, for most thermodynamic properties it is enough to know
the low energy scattering phase shifts of the particles,
which could be taken from experiment, avoiding
any model dependence.
However, if one is interested in the
chiral symmetry restoration, whose order parameters are the
quark condensates, defined as derivatives of the 
pressure with respect to the quark masses, one
needs a theoretical description of the mass dependence 
of the scattering amplitudes. 

With that aim  we turn to
Chiral Perturbation Theory (ChPT) \cite{Weinberg,chpt1,GL3},
which
provides a systematic description of 
low energy hadronic interactions. 
After identifying the pions, kaons
and the eta as the Goldstone Bosons of the QCD spontaneous
chiral symmetry breaking (pseudo Goldstone bosons, due to the
small light-quark masses), the ChPT Lagrangian is built as
the most general 
derivative and mass expansion, over $4\pi F\simeq 1.2\, \hbox{GeV}$,
(the symmetry breaking scale) respecting the
symmetry constraints. At one loop any calculation can be renormalized
in terms of just a set of parameters, $L_k(\mu)$,
 $H_k(\mu)$, ($\mu$ being the renormalization scale)
which can be determined from a few experiments 
and used for further predictions at low temperatures.

In particular, within SU(2) ChPT 
the leading
coefficients of the low temperature expansion for the pressure 
and the quark condensate 
$\langle\bar{q}q\rangle=\langle\bar{u}u+\bar{d}d\rangle$
in a hadron gas whose only interacting hadrons were the pions 
and the other hadrons were free, have
been calculated \cite{Gerber} obtaining also
an estimate, $T\simeq190\,$MeV,
of the phase transition critical temperature.  
It was also shown that the perturbative
calculation of the pressure was analogous
to the second order virial expansion when the interacting
part of the second virial coefficient is obtained from
 the one loop $\pi\pi$ ChPT scattering lengths. Other works
have studied the applicability of the virial expansion with
a chemical potential \cite{nos}, 
or the critical temperature in generalized
scenarios \cite{GChPT}.

The SU(2) approach is limited by the absence of
other interacting particles, like kaons, whose
densities at $T\simeq150$ MeV are significant \cite{Gerber}. Let us recall that
the chiral phase transition can be different 
for the SU(2) and SU(3) cases, and that
several QCD inequalities and lattice results suggest a 
chiral condensate temperature suppression 
with an increasing number of light flavors \cite{flavors},
that is, entropy and disorder. 
This effect  would 
lower the chiral critical temperature down by roughly
$20\;$ MeV \cite{lattice} (in the chiral limit).
Note that none of these results have been obtained from the
hadronic phase.

In addition, within SU(2) it is not possible to study the 
$\langle\bar{s}s\rangle$ condensate. Furthermore, in the 
chiral phase diagram, 
quark masses play  the same role as magnetic fields
in ferromagnets: Intuitively, we need 
a higher temperature
to disorder a ferromagnet when there is a magnetic field
aligned along the direction of the magnetization.
Analogously, it has been found that the SU(2) chiral condensate
melts at a lower temperature in the massless limit \cite{Gerber}.
Hence, since the strange quark mass $m_s >m_u, m_d$, 
we expect a ``ferromagnetic'' effect and a slower
temperature evolution for  $\langle\bar{s}s\rangle$ than for
 $\langle\bar{q}q\rangle$. Using a QCD-motivated effective lagrangian in the
large $N_c$ limit, this effect has been studied in \cite{hatsuda}

In this work we will use SU(3) ChPT
together with the virial expansion to estimate the
effect on $\langle\bar{q}q\rangle$ of
pions interacting with kaons and etas, and study the effect
of the different quark masses. In particular, we will obtain
the temperature dependence of $\langle\bar{s}s\rangle$
in the hadronic phase.


Thus, the second order relativistic virial expansion of the pressure
for   an inhomogeneous gas made of
three species: $i=\pi,K,\eta$ is \cite{Dashen,virial2}:
\begin{equation}
\beta P=\sum_i B_{i}(T)\xi_i + 
\sum_i\left( B_{ii}\xi_i^2 + \frac{1}{2}\sum_{j\neq i}
B_{ij}\xi_i\xi_j
\right)...,
\label{virialexp}
\end{equation}
where $\beta=1/T$ and $\xi_i=\exp({-\beta m_i})$. Expanding
up to the second order in $\xi_i$ means that we consider
only binary interactions. For a free boson gas, 
$B_{ij}^{(0)}=0$ for $i\neq j$, whereas:
\begin{eqnarray}
B_i^{(0)}&=& \frac{g_i}{2\pi^2}\int_0^{\infty} dp\, 
p^2 e^{ -\beta (\sqrt{p^2+m_i^2}- m_i)},\\
B_{ii}^{(0)}&=&\frac{g_i}{4\pi^2}\int_0^{\infty} dp\, 
p^2 e^{ -2\beta (\sqrt{p^2+m_i^2} -m_i)},
\end{eqnarray} 
and the degeneracy is $g_i=3,4,1$ for $\pi, K, \eta$, respectively.
The interactions appear through \cite{Dashen,Gerber}:
\begin{equation}
B_{ij}^{(int)}=\frac{\xi_i^{-1}\xi_j^{-1}}{2\,\pi^3}\int_{m_i+m_j}^{\infty} dE\, E^2 K_1(E/T) 
\Delta^{ij}(E),
\label{2vircoef}
\end{equation}
where $K_1$ is the first modified Bessel function
and:
\begin{equation}
\Delta^{ij}=\sum_{I,J,S} (2I+1)(2J+1)\delta^{ij}_{I,J,S}(E),
\label{Delta}
\end{equation}
$\delta^{ij}_{I,J,S}$ being the $ij\rightarrow ij$
phase shifts (chosen so that $\delta=0$ at threshold) 
of the  elastic scattering of a state
$ij$ with quantum numbers $I,J,S$ ($J$ being the total angular momentum
and $S$ the strangeness).

The virial expansion seems to break \cite{nos} at 
$T\simeq 200-250$ MeV. Fortunately,
the physics we are interested in occurs already at  $T<250\,$MeV,
and $\xi_\pi> \xi_K \simeq \xi_\eta$, so that it is enough
to consider $ij=\pi\pi,\pi K$ and $\pi \eta$ in the second virial
coefficients. Thus we are considering up to $\xi_\pi$ suppressed
corrections to the $\pi, K,\eta$ dilute free gas.

Let us recall that the pressure is 
nothing but the $T$ dependent part 
of the free energy density $z$, i.e., $P=\epsilon_0- z$,
$\epsilon_0$ being the  vacuum energy density.
Hence, \cite{Gerber}:
\begin{equation}
\langle\bar{q}_\alpha q_\alpha\rangle=\frac{\partial z}{\partial m_{q_\alpha}}
=\langle 0 \vert\bar{q}_\alpha q_\alpha\vert 0 \rangle-  
\frac{\partial P}{\partial m_{q_\alpha}}.
\label{condmq}
\end{equation}
where $q_\alpha=u, d,s$.
Let us emphasize that,
in contrast with most thermodynamic quantities, for
the chiral condensate it is not enough
to know the $\delta(E)$, but {\it we also need
their dependence with the quark masses} as well as a value
for the vacuum expectation value. All that will  be obtained from
ChPT. For that purpose, we first
rewrite the condensate, eq.(\ref{condmq}), in terms of meson masses:
\begin{equation}
\langle\bar{q}_\alpha q_\alpha\rangle
=\langle 0 \vert\bar{q}_\alpha q_\alpha\vert 0 \rangle 
\left(1+\sum_i \frac{c^{\bar{q_\alpha}q_\alpha}_i}{2 m_i F^2} 
\frac{\partial P}{\partial m_i}\right)
\end{equation}
where, as before, $i=\pi,K, \eta$. Note that we have defined:
\begin{equation}
c^{\bar{q_\alpha}q_\alpha}_i=- F^2\frac{\partial m_i^2}
{\partial m_{q_\alpha}}\langle 0 
\vert\bar{q}_\alpha q_\alpha\vert 0 \rangle ^{-1}.
\end{equation}
In the isospin limit,
when $m_u=m_d$, the $u$ and $d$ condensates
are equal, and we define
$\langle 0\vert\bar{q} q\vert 0 \rangle \equiv\langle 0\vert\bar{u} u+\bar{d} d\vert 0 \rangle$.
It is tedious but straightforward to obtain expressions 
for the $c$ parameters above. They can be found in the appendix.
Both the $\langle 0\vert\bar{q} q\vert 0 \rangle$ and 
$\langle 0\vert\bar{s} s\vert 0 \rangle $ one loop calculations 
within SU(3) ChPT were given in \cite{GL3}, 
together with the one loop dependence of the meson masses 
on the quark masses needed to obtain 
$\partial m^2_i/\partial m_{q_\alpha}$.
The only relevant comment is that the $c$ coefficients depend on
the chiral parameters $L_k$, for $k=4...8$, and $H_2$.

In Table I, we show several $L_k$ determinations
from meson data. It makes little difference to use
other determinations. For $H_2$ we will
use $H_2^r(M_\rho)=(-3.4\pm1.1)10^{-3}$, obtained 
as explained in \cite{Jamin} but using a more
recent estimation of 
$\langle 0\vert\bar{s}s\vert 0\rangle/\langle 0\vert\bar{q}q\vert 0\rangle=0.75\pm 0.12$
\cite{Narison}.
For instance,  the
$c$ parameters obtained when using the $L_k$ in the first
column of Table I are:
\begin{eqnarray}
c^{\bar{q}q}_\pi=0.9^{+0.2}_{-0.4} , c^{\bar{q}q}_K=0.5^{+0.4}_{-0.7},
c^{\bar{q}q}_\eta=0.4^{+0.5}_{-0.7}, \nonumber\\
c^{\bar{s}s}_\pi=-0.005^{+0.029}_{-0.037}, c^{\bar{s}s}_K=1.3^{+0.4}_{-0.8},
c^{\bar{s}s}_\eta=1.5^{+0.9}_{-1.6}. \nonumber
\end{eqnarray}
Note that our $c^{\bar{q}q}_\pi$ is in a good
agreement with the SU(2) estimates: $0.85$ and  $0.90\pm0.05$ \cite{Gerber}.
Had we used the 
estimate $H_2^r(M_\rho)\simeq2L_8^r(M_\rho)\simeq1.8\times10^{-3}$
from scalar resonance saturation \cite{saturation},
all our following results would remain unchanged for $\langle\bar{q}q\rangle$,
but  the strange condensate {\it would not even seem to melt} within this
approximation.
\begin{table}
\begin{ruledtabular}
\begin{tabular}{|c||c|c|c|}
&Refs.\cite{chpt1,BijnensGasser}
&Ref.\cite{BijnensKl4}
&IAM \cite{angelyo}
\\ \hline
$L_1^r(M_\rho)$
& $0.4\pm0.3$
& $0.46$
&$0.561\pm0.008\,(\pm0.10)$\\
$L_2^r(M_\rho)$
& $1.35\pm0.3$
& $1.49$
&$1.21\pm0.001\,(\pm0.10)$\\
$L_3$
& $-3.5\pm1.1$
& $-3.18$
&$-2.79\pm0.02\,(\pm0.12)$\\
$L_4^r(M_\rho)$
& $-0.3\pm0.5$
& 0
&$-0.36\pm0.02\,(\pm0.17)$\\
$L_5^r(M_\rho)$
& $1.4\pm0.5$
& $1.46$
&$1.4\pm0.02\,(\pm0.5)$\\
$L_6^r(M_\rho)$
& $-0.2\pm0.3$
& 0
&$0.07\pm0.03\,(\pm0.08)$\\
$L_7$
& $-0.4\pm0.2$ 
& $-0.49$
&$-0.44\pm0.003\,(\pm0.15)$\\
$L_8^r(M_\rho)$
& $0.9\pm0.3$ 
& $1.00$
&$0.78\pm0.02\,(\pm0.18)$\\
\end{tabular}
\end{ruledtabular}
\caption{\rm Different sets of chiral parameters $\times10^{3}$. 
In the second column $L^r_1,L^r_2,L_3$ are taken from
\cite{BijnensGasser}
and  the rest from  \cite{chpt1}
($L^r_4$ and $L^r_6$ are estimated from the Zweig rule).
The third column comes from an  $O(p^4)$
analysis of $K_{l4}$ decays \cite{BijnensKl4} 
($L^r_4$ and $L^4_6$ are set to zero).
The last column is the IAM fit \cite{angelyo}
to the whole meson-meson scattering up to 1.2 GeV.
}
\label{eleschpt}
\end{table}
Finally, we need the meson-meson amplitudes
to one-loop in SU(3) ChPT, which have been given in \cite{o4,angelyo}.
These expressions have to be projected in partial waves
of definite isospin $I$ and angular momentum $J$, whose complex phases
are the $\delta_{I,J}(E)$ in eq.(\ref{Delta}).
ChPT is known to provide  a good low energy description ($E<500\,$MeV)
of the meson-meson amplitudes \cite{o4}.
Within SU(2) ChPT it has been shown \cite{Gerber}
 that using just the amplitudes at threshold 
(scattering lengths) 
in the virial expansion is equivalent to expanding the partition
function to third order in $T/F$ or $T/m_\pi$. This approach yields
a fairly good representation of the pion gas thermodynamics
at low temperatures, $T< 150\,$MeV \cite{Gerber}.

The virial coefficients and 
$\partial P/\partial m_i$ have been calculated numerically.
For the figures, we have represented the 
chiral condensate over its vacuum expectation value,
$\langle \bar{q}_\alpha q_\alpha
\rangle /\langle 0 \vert\bar{q}_\alpha q_\alpha\vert 0 \rangle $,
so that all of them are normalized to 1 at $T=0$.
In reality, since there is always an small explicit chiral symmetry 
breaking due to the quark masses, the condensate does not 
vanish completely, except in the $T\rightarrow\infty$ limit
(following with our analogy, our ferromagnet above
the Curie point becomes paramagnetic in the presence of a magnetic field).
Of course, with the second order virial approach we cannot 
generate such an analytic behavior, and our curves become negative
above some $T$. Since the non-strange quark
masses are very tiny compared with the  $\langle \bar{q} q\rangle$
(about a 3\%), it is a good approximation to
let the
condensate vanish and talk about a melting temperature.
This is not the case for the strange condensate, which
has a larger explicit breaking due to the much
larger mass of this quark. 

In Figure 1 we show the results of using the 
one-loop ChPT expansion for the phase shifts.
The thin continuous line represents the interacting pion gas in SU(2),
but updated to the SU(2) parameters obtained from 
those in the first column
of Table I (see \cite{GL3} for the conversion), whereas
the thin-dashed line is the corresponding
one-loop SU(3) result also for a pion gas. We have checked that
the tiny difference between them comes only from the $O(p^4)$
phase shift contribution, ( it is
 the same for SU(2) and SU(3) only up to $s/M_K^2$ or $s/M_\eta^2$
terms from expanding kaon or eta loops \cite{GL3}). 
This amounts to a  4 MeV decrease
of the pion gas extrapolated melting temperature: 
$T_m^{\langle \bar{q} q\rangle}=231\,$MeV for SU(2) and 
$T_m^{\langle \bar{q} q\rangle}=227\,$MeV for SU(3). Next, the thin-dotted line is obtained 
by adding free kaons and etas to the pion gas,
which lowers the extrapolated melting temperature down to 221 MeV.

Our new SU(3) result for a 
$\pi, K, \eta$ gas, is the thick continuous line,
where we see that the additional decrease due to
$\pi K,\pi\eta$ interactions amounts to 10-11 MeV.
It may seem surprising that the 
$\pi K, \pi\eta$ interaction effect may be comparable
or larger than that of  free kaons and etas, since they
are thermally suppressed. Note, however, that
$\exp(-m_\pi/T)$ just amounts to a factor of 6,4,2.5 
suppression at T= 80,100,150 MeV, respectively.
In contrast, the $\pi K$ 
and $\pi\eta$ interactions depend strongly on $m_\pi$, much more
sensitive to $\hat m$ than  $m_K$ or $m_\eta$, which are the
only dependence of the free $K$ and $\eta$ terms on $\hat m$.
In particular 
$\partial m_\pi^2/\partial \hat m\simeq 2 \, \partial m_K^2/\partial\hat m$
and
$\partial m_\pi/\partial \hat m\simeq 6 \, \partial m_K/\partial\hat m$, 
and this ``temperature independent enhancement'' 
competes with the thermal suppression, making the $\pi K$ and $\pi \eta$
interaction effect comparable to the free one around  
$T=80-100\,$MeV and larger if $T>140\,$MeV.
All together, the $T_m^{\langle \bar{q} q\rangle}$ decrease in SU(3) is
roughly 20 MeV, in agreement 
with the chiral limit lattice 
results $T_c=173\,$MeV for SU(2) and  $T_c=154\,$MeV,
for SU(3) \cite{lattice}.

Furthermore,
 it can be noticed that $\langle \bar{s} s\rangle$ melts much slower
than  $\langle \bar{q} q\rangle$,
since $m_s\gg \hat m$. Indeed, there is still 70\% left of
the $\langle \bar{s} s\rangle$ condensate at the  $\langle \bar{q} q\rangle$
melting point. 
Let us recall that both effects are already sizable at 
low temperatures $T\simeq100\, $MeV, even lower
when looking at the slower $\langle \bar{s} s\rangle$ evolution,
which already separates from the $\langle \bar{q} q\rangle$
evolution at temperatures as low as $70\,$ MeV.
Since we can still trust 
pure ChPT up to, roughly, $T\simeq150\,$MeV \cite{Gerber},
we conclude that both effects are model independent predictions.

However, just for illustration and to ease 
the comparison between curves and with previous works,
we have extrapolated the condensates to higher temperatures. 
We have also made a Montecarlo Gaussian sampling of the chiral
parameters within their error bands. We find that the melting
temperatures would be:
\begin{eqnarray*}
&&T_m^{\langle \bar{q} q\rangle SU(3)}
= 211^{+19}_{- 7}\,\hbox{MeV}, \\
&&T_m^{\langle \bar{q} q\rangle SU(2)}
= 231^{+30}_{-10}\,\hbox{MeV}\\
&&T_m^{\langle \bar{q} q\rangle SU(2)}-T_m^{\langle \bar{q} q\rangle SU(3)}
= 21^{+14}_{-7}\,\hbox{MeV}
\end{eqnarray*}
Note that the two melting temperatures 
are strongly correlated.

Concerning the $\langle \bar{s} s\rangle$ extrapolation, its
``melting'' temperature 
$T_m^{\langle \bar{s} s\rangle}= 291^{+37}_{-35}\,\hbox{MeV}$,
is just for  illustration, since, apart from
lying beyond the reliability region of the approach, as stated above,
actually it should not melt completely, except in the $T\rightarrow\infty$
limit.
In particular, we would find
$T_m^{\langle \bar{s}s\rangle}-T_m^{\langle \bar{q}q\rangle}
= 80^{+25}_{-40}\,\hbox{MeV}$.
Nevertheless, we remark once more that
the slower $\langle \bar{s} s\rangle$
thermal evolution is clearly seen at temperatures where the
whole approach can be trusted. As a matter of fact, 
from the figure, we see that the
strange condensate does not start to melt sizably
up to $T\simeq150\,$MeV.

\begin{figure}[h]
\includegraphics[scale=.51]{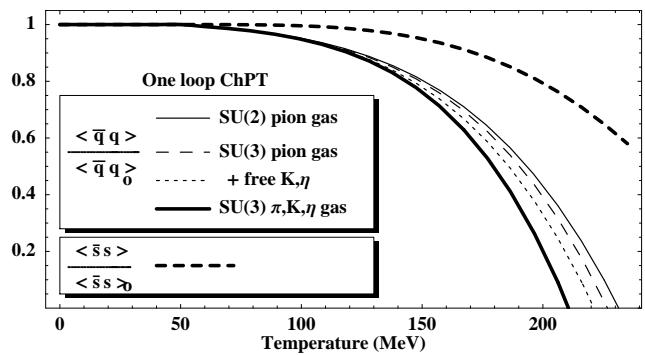}
\caption{\rm Condensate evolution with temperature using one-loop ChPT
amplitudes. Although {\it they should not really vanish}
they are extrapolated down to zero only for reference.}
\end{figure}

Furthermore, we have also estimated the effect of other, 
more massive, hadrons, which in the SU(2) case \cite{Gerber,nos}
also included the kaons and etas,
and reduced $T_m^{\langle \bar{q}q\rangle}$ by approximately
10-20 MeV. Since they are heavier than $m_\eta$, their
density is very low, and their main contribution to the pressure
comes from the first virial coefficient, i.e. the free gas.
The only uncertainty is on $\partial M_h/\partial m_{q_\alpha}$, 
conservatively estimated to lie within
the number of valence quarks $N_{q_\alpha}$ and 
$2N_{q_\alpha}$. Thus, we have found that their 
contribution decreases $T_m^{\langle \bar{q} q\rangle}$
by 7-12 MeV, respectively.
The heavy hadron contribution is the same for the SU(2) and SU(3) cases,
so that the difference between their respective melting temperatures
 remains the same.
Finally if we were to extrapolate the strange condensate
evolution we would also find a decrease of 
$T_m^{\langle \bar{s} s\rangle}$ by 15-23 MeV due to
these heavier hadrons.

As an attempt to estimate the uncertainty due to the 
truncation of the virial expansion, we have added 
to our calculation the third order coefficient for free pions,
which can be calculated exactly. If we do so, the melting temperature
decreases by one MeV. It seems that when higher orders corrections 
start becoming important, 
 the curve is already so steep that the melting point
changes very slightly.
We cannot calculate the size of the 
three pion interaction term, but we expect it to be of the same order, 
and consistently with what we did for the heavy hadron uncertainty,
we also allow for twice as much uncertainty, i.e., another 2 MeV. 
The important point, however,  is that the error is 
dominated by the uncertainty 
in the chiral parameters, which is an order of magnitude larger. 
Altogether, we find:
\begin{equation}
T_m^{\langle \bar{q}q\rangle}\simeq 201^{+23}_{-11}\,\hbox{MeV}
\end{equation}

Finally, and in order to estimate the effects of higher energies
we can extend ChPT
up to $E\simeq1.2\,$GeV by means of unitarization models
\cite{IAM1,IAM2,angelyo}.
These techniques resum the ChPT series respecting unitarity but
also the low energy expansion, including the mass dependency.
In particular, it has been shown that the coupled channel
Inverse Amplitude Method (IAM) provides a remarkable and accurate description
of the complete meson-meson interactions 
 below 1.2 GeV, generating dynamically
six resonances  from
the one-loop ChPT expansion and a set of fitted $L_k$ compatible
with other previous determinations. This approach can be 
extended systematically to higher orders 
(for SU(2) case, see \cite{Juan}).

Hence, in Figure 2 we show the results of using the
one-loop coupled channel IAM fitted phase shifts. Its
corresponding $L_i$ parameters are given in
the last column of Table I, with two errors, the first, very small, is
purely statistical, and the second covers the uncertainty in the
parameters depending on what systematic error is assumed for 
the experimental data. Let us remark that, although 
 the $L_k$ are highly correlated, the second, larger error, 
ignores completely these correlations, 
and should be considered as a very conservative range.   
The continuous line corresponds to the central values, 
and the dark shaded areas cover the one standard deviation uncertainty
due to the small errors in the parameters.
These areas have been obtained from a Montecarlo Gaussian sampling.
The conservative ranges are covered by the light gray areas.
We now find:
\begin{eqnarray*}
&&T_m^{\langle \bar{q} q\rangle SU(3)}
= 204^{+3}_{-1}\,\,(^{13}_{\,\,5})\, \hbox{MeV},\\
&&T_m^{\langle \bar{q} q\rangle SU(2)}
= 235^{+3}_{-1}\,\,(^{14}_{\,\,1})\, \hbox{MeV},\\
&&T_m^{\langle \bar{q} q\rangle SU(2)}
-T_m^{\langle \bar{q} q\rangle SU(3)}=
31.50^{+1.20}_{-0.03}\,\,(^{9}_{8})\, \hbox{MeV}
\end{eqnarray*}
where the errors in parenthesis are the conservative
errors which should be interpreted
as uncertainty ranges better than as  standard deviations. 
Note the excellent agreement with standard ChPT. 
The magnitude of the different contributions is roughly the same,
although the $\pi K$ and $\pi\eta$ interactions this time lower
$T_m^{\langle \bar{q} q\rangle}$ by 17 MeV, 
and the free kaons and etas by roughly 10 MeV.
Let us remark that the strange condensate again melts
much slower than the non-strange condensate, and that this effect
is clearly sizable at low temperatures, where both ChPT
and the virial expansion can be trusted.
With all the caveats presented above, and just for illustration
we also extrapolate $\langle \bar{s} s\rangle$ to zero
and we find: $T_m^{\langle \bar{s} s\rangle}
= 304^{+39}_{-25}\,\,(^{120}_{\,\,65})\, \hbox{MeV}$,
in particular, the strange condensate melting
is retarded by $T_m^{\langle \bar{s}s\rangle}-
T_m^{\langle \bar{q}q\rangle}= 
100^{+36}_{-29}\,\,(^{120}_{\,\,80})\,\hbox{MeV}$.

\begin{figure}[h]
\includegraphics[scale=.6]{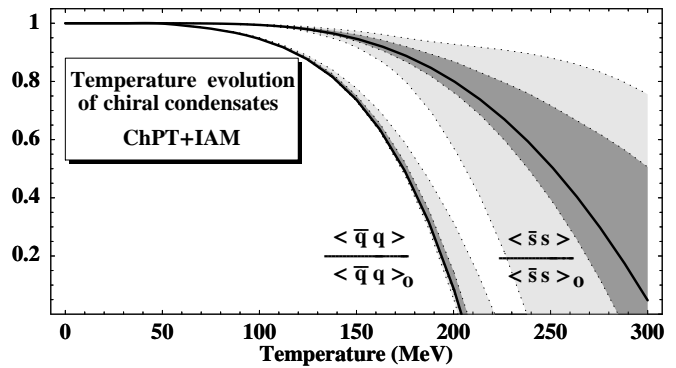}
\caption{\label{fig:epsart} \rm Temperature
evolution of chiral condensates using the unitarized ChPT amplitudes. 
The shaded areas cover the uncertainties
in different sets of chiral parameters. Although 
{\it they should not really vanish}
they are extrapolated down to zero only to ease their comparison 
and for reference.
}
\end{figure}
Again other massive states apart from kaons and etas
would lower $T_m^{\langle \bar{q} q\rangle}$
by 6-10 MeV, and $T_m^{\langle \bar{s} s\rangle}$ by 22-32 MeV.
Also very similar is the effect of including the third order
virial coefficient for free pions. As it happened in the
standard ChPT case, it decreases the melting temperature by one MeV,
and we add another 2 MeV of uncertainty for the unknown interacting part, that is
\begin{equation}
T_m^{\langle \bar{q} q\rangle}\simeq 195 ^{+8}_{-5}\,\,(^{17}_{\,7})\, \hbox{MeV}
\end{equation}
Again, the errors correspond to the uncertainty
in the chiral parameters plus that on the heavier states,
plus that from the truncation of the virial expansion
and those in parenthesis should be interpreted as conservative ranges.

These results show that higher energy effects do not
affect very much our conclusions and that the 
results of the integrals in eq.(\ref{2vircoef}),
which extend to infinity, seem robust.

We have studied the SU(2) and SU(3) 
temperature evolution of the chiral condensates {\it in the
hadronic phase}. The thermodynamics of the meson gas has been 
obtained from the virial expansion and Chiral Perturbation Theory.
Our results clearly show a significant 
decrease, about 20 MeV,  of the non-strange condensate melting temperature,
from the SU(2) to the SU(3) case, similar to lattice results.
 Of these, about 6 MeV
had already been explained with free kaons and etas, but
the rest are mainly due to $\pi K$ and $\pi \eta$ interactions.
In addition, our results show an slower 
temperature evolution of the strange condensate, 
about 90\% 
of its zero value still remaining at 150 MeV, 
due to the different quark masses.
Both effects are clearly visible already at low temperatures.
However, we have checked that
their size is completely similar when using unitarized models.

These techniques should be easily extended 
to Heavy Baryon Chiral Perturbation Theory, in order to study
the condensates with non-zero baryon density. In general the whole
approach could be used with any effective Lagrangian formalism,
and in particular to study other QCD phase transitions like those of the
color superconducting phases.

\paragraph{Appendix.-}
We provide here the expressions for
the SU(3) $c_i$ coefficients:
\begin{eqnarray*}
c^{\bar{q}q}_\pi&&  =1+4\mu_\pi+2\mu_K+m_\pi^2(\nu_\pi-\frac{\nu_K}{9})-\frac{16 m_K^2}{F_0^2}L_4\\
&&+\frac{4 m_\pi^2}{F_0^2}(8L_6+6L_8-6L_4-4L_5-H_2)
\end{eqnarray*}
\begin{eqnarray*}
c^{\bar{q}q}_K&& =\frac{1}{2}+\frac{3\mu_\pi}{2}+\mu_K+\frac{\mu_\eta}{2}
-\frac{4 m_\pi^2}{F_0^2}\left(L_4+L_8+\frac{H_2}{2}\right)
\\
&&+\frac{2 m_K^2\nu_\eta}{9}+\frac{8 m_K^2}{F_0^2}
(4 L_6+2L_8-3L_4-L_5)
\end{eqnarray*}
\begin{eqnarray*}
  c^{\bar{q}q}_\eta&&=\frac{1}{3}\left\{
1+6\mu_K-3m_\pi^2\nu_\pi+\frac{\nu_\eta}{3}(m_\pi^2-4m_\eta^2)\right.\\
&& +\nu_K(3m_\eta^2+m_\pi^2)+\frac{16m_\eta^2}{F_0^2}(6L_6+2L_8-L_5-3L_4)\\
&&-\frac{16L_4}{F_0^2}(m_K^2+m_\pi^2/2)
-\frac{128}{3F_0^2}(m_K^2-m_\pi^2)(3L_7+L_8)\\
&&-\frac{4m_\pi^2}{F_0^2}(2L_8+H_2)\}
\end{eqnarray*}
\begin{eqnarray*}
c^{\bar{s}s}_\pi&& =\frac{16m_\pi^2}{F_0^2}(2L_6-L_4)-\frac{4}{9}\nu_\eta
m_\pi^2
\end{eqnarray*}
\begin{eqnarray*}
c^{\bar{s}s}_K&& =
\{ 1+2\mu_\eta+4\mu_K+\frac{4m_\pi^2}{F_0^2}[2(L_8-L_4)+H_2]\\
&&
+\frac{8}{9}\nu_\eta m_K^2+\frac{16m_K^2}{F_0^2}
(2L_6+L_8-2L_4-L_5-H_2/2)\}
\end{eqnarray*}
\begin{eqnarray*}
c^{\bar{s}s}_\eta&& = \frac{4}{3}\{1+6\mu_K+m_\eta^2(3\nu_K/2-4\nu_\eta/3)\\
&&+m_\pi^2(\nu_K/2+\nu_\eta/3)
+\frac{16m_K^2}{F_0^2}\left(4L_7-L_4-\frac{H_2}{2}+\frac{L_8}{3}\right)\\
&&
+\frac{8m_\pi^2}{F_0^2}\left(-L_4-\frac{5}{3}L_8+\frac{H_2}{2}
-8L_7\right)\\
&&
+\frac{8m_\eta^2}{F_0^2}\left(4 L_8-2L_5-\frac{3L_4}{2}
+3L_6\right)
\end{eqnarray*}
where following the standard notation in ref.\cite{GL3} 
$\mu_i=m_i^2\log(m_i^2/\mu^2)/(32\pi^2F_0^2)$,
$\nu=(\log(m_i^2/\mu^2)+1)/(32\pi^2F_0^2)$ and $F_0$ is 
the pion decay constant in  the SU(3) chiral limit.
In this work we have set the renormalization scale to $\mu=M_\rho$

 The author
thanks  A. Dobado, A. G\'omez Nicola and E. Oset 
for useful comments,
and support from the Spanish CICYT projects
FPA2000 0956 and BFM2000 1326, as well as a
Marie Curie fellowship MCFI-2001-01155 .

\bibliography{apssamp}


\end{document}